# Evidence for cascaded third harmonic generation in non-centrosymmetric gold nanoantennas

Michele Celebrano,[1]‡* Andrea Locatelli,[2]‡ Lavinia Ghirardini,[1] Giovanni Pellegrini,[1] Paolo Biagioni,[1] Attilio Zilli,[1] Xiaofei Wu,[3,4] Swen Grossmann,[3,4] Luca Carletti,[2] Costantino De Angelis,[2] Lamberto Duò,[1] Bert Hecht,[3,4] Marco Finazzi.[1]

[1] Physics Department, Politecnico di Milano, Piazza Leonardo Da Vinci 32, 20133 Milano, Italy.

[2] Department of Information Engineering, University of Brescia, Via Branze 38, 25123 Brescia, Italy.

[3] Nano-Optics & Biophotonics Group - Department of Physics - Experimental Physics 5, University of Würzburg, Am Hubland, 97074 Würzburg, Germany.

[4] Röntgen Research Center for Complex Material Systems (RCCM), Am Hubland, 97074 Würzburg, Germany.

ABSTRACT

The optimization of nonlinear optical processes at the nanoscale is a crucial step for the integration of complex functionalities into compact photonic devices and metasurfaces. In such systems, photon upconversion can be achieved with high efficiencies via third-order processes, such as third harmonic generation (THG), thanks to the resonantly enhanced volume currents. Conversely, second-order processes, such as second harmonic generation (SHG), are often inhibited by the symmetry of metal lattices and of common nanoantenna geometries. SHG and THG processes in plasmonic nanostructures are generally treated independently, since they typically represent small perturbations in the light-matter interaction mechanisms. In this work, we demonstrate that this paradigm does not hold for plasmon-enhanced nonlinear optics, by providing evidence of a sum frequency generation process seeded by SHG, which sizably contributes to the overall THG yield. We address this mechanism by unveiling a characteristic fingerprint in the polarization state of the THG emission from non-centrosymmetric gold nanoantennas, which directly reflects the asymmetric distribution of second harmonic fields within the structure and does not depend on the model one employ to describe photon upconversion. We suggest that such cascaded processes may also appear for structures that exhibit only moderate SHG yields. The presence of this peculiar mechanism in THG from plasmonic nanoantennas at telecommunication wavelengths allows gaining further insight on the physics of plasmon-enhanced nonlinear optical processes. This could be crucial in the realization of nanoscale elements for photon conversion and manipulation operating at room-temperature.

TEXT

Since the first experimental evidences of enhanced nonlinear optical effects in nanostructured metals,[1,2] the field of nonlinear plasmonics has been experiencing a continuous growth[3-6] aimed at integrating multiple optical functionalities into ultra-compact all-optical devices.[7] Highly efficient nonlinear photon conversion in extremely confined volumes constitutes, indeed, a crucial step towards the realization of nanoscale photon conversion units[8] for quantum networks working at room temperature. In this frame, plasmonic nanoantennas can effectively compensate for the lack of phase-matching conditions[9] at the nanoscale by means of intense field enhancements and of the steep field gradients generated by the antenna hotspots at the plasmon resonances.[10,11]

One of the most efficient upconversion mechanisms in plasmonics, third harmonic generation (THG), consists in the absorption of three photons with equal energy followed by the emission of a fourth photon at three times the energy of the impinging ones. For plasmonic nanoantennas working in the visible range, the emission yield of THG is commonly reported to be orders of magnitude higher than that of Second Harmonic Generation (SHG), which entails the emission of a photon at twice the energy of the two impinging ones. In fact, THG benefits from the resonantly-enhanced volume currents in plasmonic nanoantennas at visible wavelengths, a process typically modeled by considering the bulk $\chi^{(3)}$ material nonlinear susceptibility.[12-16] On the contrary, SHG is strictly forbidden in the bulk of centrosymmetric materials, such as those usually employed in plasmonics (Au, Ag, Al), where SHG is due to surface-related processes stemming from the removal of inversion symmetry associated with the metal/environment interface[17-24] or due to strong field gradients and nonlocal effects.[17,18] Such SHG mechanisms are typically described by an effective surface $\chi^{(2)}$ nonlinear susceptibility,

The mechanisms at the basis of THG in metal nanostructures are still largely debated,[25-27] since the driving fields are strongly nonuniform within the skin depth inside the material and retardation effects might play a fundamental role. For this reason, the values of $\chi^{(3)}$ extrapolated from plasmonic nanostructures display a large variability, depending on the sample quality in terms of surface roughness and/or on the spectral position of the plasmon resonances with respect to the driving fundamental wavelength (FW) and third harmonic (TH) frequencies.[28-31] In particular, state-of-the-art modelling seems to fail in predicting the behavior of $\chi^{(3)}$ in the near-infrared region,[25-27] where applications to sensing can profit from the low absorption of biological entities and nonlinear photon conversion would become compatible with optical fiber technology and silicon-based photonics. According to largely-adopted theoretical models, a significant drop of $\chi^{(3)}$ is expected in the near infrared since, neglecting the frequency dependence of the nonlinear susceptibility, the third-harmonic intensity should scale as the fourth power of the inverse wavelength.[30] Although this picture was confirmed by some recent trials,[32] other experimental studies report extremely large THG conversion efficiencies in the same spectral range for both plasmonic[12] and hybrid nanoantennas.[33,34] This indicates third-order nonlinear processes as a very promising route towards efficient photon conversion at the nanoscale even for infrared photons. Further investigations are therefore essential to gain more insight into the processes promoting plasmon-enhanced THG in this wavelength range.

In this paper, we provide experimental evidence for the presence of a two-step cascaded process in the THG from non-centrosymmetric gold nanoantennas,[35] which adds up to the THG originating from a direct upconversion of three FW photons. We ascribe this phenomenon to the coherent built-up of sum frequency generation (SFG) process occurring between a FW pump photon at $\omega$ and a second harmonic (SH) photon at $2\omega$, which is seeded by a surface $\chi^{(2)}$-mediated process. As predicted by full-wave simulations, the latter leaves a unique fingerprint in the THG polarization state, which deviates from that of a purely $\chi^{(3)}$-mediated effect. We unequivocally identify such behavior through the detection and analysis of the polarization states of both SHG and THG from the nanoantennas.

We investigate nanostructures realized following the same procedure as described in Ref. 35 and depicted in the inset of Fig. 1. The antennas are engineered to feature a V-shaped nanostructure that determines the lowest-order antenna resonance, coupled to a nanorod whose function is to shift spectral weight across the $2\omega$ frequency by plasmon hybridization. As in Ref. 35, the aim is to tune the first and second resonance of the structure to be degenerate with the pump frequency $\omega$ and with $2\omega$, respectively, for efficient SHG. We focus our attention on a set of antennas having the same geometry and size of the V-shaped nanostructure, corresponding to a first resonance at the FW wavelength (1554 nm) and resulting in a comparable total antenna volume, a choice that provides similar conditions for the bulk contribution to THG. Yet, this set of antennas is characterized by rods having different lengths (or by no rod at all), providing different resonance conditions for the structure at the SH.

We characterize and compare the nonlinear emission of the nanostructures using the same microscopy setup employed in Ref. 35. The ultrafast pump pulses, emitted by an Er-doped fiber laser are linearly polarized (polarization ratio $10^3$:1) to match the fundamental mode of the antenna and then focused by a 0.85 numerical aperture (NA) air objective, resulting in a pump intensity ranging from 0.04 to 0.2 GW/cm$^2$ in the focal spot. The nonlinear emission is collected by the same objective and, after a nonpolarizing beam-splitter and a filter cutting the residual pump radiation, is sent to either a visible/near infrared (Vis-NIR) spectrometer or a single photon avalanche detector (SPAD). Possible nonlinear emission from the substrate remains below the detection limit in the whole employed pump intensity range. The polarization state of the emitted photons is analyzed by inserting a rotating broadband polarizer (extinction ratio ≈ $10^4$:1) in the collection path before the detectors.

A complete set of data presenting the SHG and THG emission properties determined on two samples with nominally equal geometrical parameters is presented in the Supplementary Information (Figs. S1 and S2). Here, we will focus our attention on the most efficient doubly resonant antenna in terms of THG and SHG (from a third sample with improved geometrical parameters), Antenna R, as displayed in Fig. 1. For comparison, we will also report THG and SHG data obtained on a second antenna, Antenna V, which has no nanorod but is characterized by the same V-shape geometry (see Fig. 1). Figure 2a reports the nonlinear emission spectrum of Antenna R, characterized by two narrow peaks around 518 nm and 777 nm that correspond to the TH and the SH radiation emitted by the nanoantenna, respectively. To single out the individual SHG and THG contributions, although the incoherent luminescence emission from these nanostructures[36] is barely detectable, we select two narrow spectral regions (see arrows in Fig. 2a) around the two emission peaks using band-pass filters. Figures 2b and 2c show the THG and SHG power curves acquired on the two antennas, R and V. A comparison between the power curves indicates that the SHG has a quadratic dependence on the pump power, while the THG has the typical cubic power dependence. As already previously reported,[35] Antenna V is less efficient than Antenna R, in terms of both THG and SHG. By considering the light collected and transmitted by the objective, the transmission of the optics and the quantum efficiency and the filling-factor of the SPAD, we estimate the effective THG (SHG) power emitted by Antenna R to be about 1.1 pW (3.9 pW) at 0.2 GW/cm² pump fluency. This allows achieving maximum conversion efficiencies $\eta_{\text{THG}} \approx 1.1 \times 10^{-8}$ and $\eta_{\text{SHG}} \approx 3.9 \times 10^{-8}$ as well as nonlinear coefficients $\gamma_{\text{THG}} = \frac{P_{\text{THG}}}{(P_{\text{FW}})^3} \approx 1.1 \text{ W}^{-2}$ and $\gamma_{\text{SHG}} = \frac{P_{\text{SHG}}}{(P_{\text{FW}})^2} \approx 3.9 \times 10^{-4}$ W$^{-1}$, where $P_{\text{THG}}$, $P_{\text{SHG}}$ and $P_{\text{FW}}$ represent the THG, SHG and FW measured peak powers, respectively.

The polar plots representing the THG and SHG intensities for the three antennas as a function of the rotation angle of the polarizer are shown in Figs. 2d and 2e, respectively. While the main polarization axis of the SHG is orthogonal to the polarization axis of the FW light (i.e. parallel to the symmetry plane of the V-shaped structure), as expected,[35] the polarization state of the THG presents some unexpected features: (i) the maximum emission is obtained when the polarizer angle is set at about 75° for antenna R, evidencing a significant tilt of the THG main polarization axis with respect to the linearly polarized pump electric field (see dark red double-headed arrow). Conversely, the THG main polarization axis of Antenna V is parallel (within the experimental uncertainty) to the FW field; (ii) the degree of linear polarization, which is related to the ratio between the maximum and the minimum of the intensity in the polar plot, is less pronounced than that of SHG, especially for Antenna R. Qualitatively similar results have been reproduced on two sets of nanostructures with nominally equal geometrical parameters, as reported in the Supplementary Information (see Figs. S2 and S3). While the SHG polarization remains perpendicular to the FW electric field for all antennas, the main polarization axis of the THG emission is observed to be parallel to the FW polarization for V-shape antennas that have no nanorod counterpart, while it is systematically rotated counterclockwise for V-shaped nanostructures coupled to nanorods. Possible defects or fabrication uncertainties introduce some variability in the tilt of the TH polarization axis and in the THG and SHG intensities for nominally identical antennas. However, the systematic behavior of the THG emission from so many antennas allows us to rule out defects or surface roughness as the possible cause of the observed tilt of the TH polarization axis.

To understand these peculiar observations, we numerically evaluated the nonlinear polarized emission of the nanoantennas supported by a perfectly linear substrate having a refractive index equal to 1.512. We performed full-wave finite-element simulations that consider the SH and TH emission of the nanostructures as the result of several nonlinear phenomena. To account for the process directly converting three photons at the FW as sketched in Fig. 3a, we considered three different mechanisms as possible driving sources for THG. The first one is a bulk $\chi^{(3)}$-mediated processes, modeled by considering gold as an isotropic material that induces a nonlinear polarization term at the third harmonic: $\mathbf{P}_{3\omega} = \chi^{(3)}(\mathbf{E}_\omega \cdot \mathbf{E}_\omega)\mathbf{E}_\omega$ (in Gauss units).[26-31] The corresponding normalized THG emission plot is reported in Fig. 3d. A second source, encompassing non-local and gradient field effects that are known to contribute for strong field confinement in the nonlinear optical response in metals,[37] is obtained by analytically solving the Euler equations for the plasma dynamics as in Ref. 38. The leading order contributions to $\mathbf{P}_{3\omega}$ depending on the FW fields inside the nanostructures were obtained by neglecting the electron pressure,[38] which is assumed to be counterbalanced by the force due to the surface-barrier potential. Finally, we considered a surface contribution to $\mathbf{P}_{3\omega}$, proportional to the cube of the $\mathbf{E}_\omega$ field at the surface. This surface $\mathbf{P}_{3\omega}$ term is assumed to be perpendicular to the antenna surface, to mimic the behavior of the leading term for surface SHG (see below). Irrespectively of the relative THG efficiencies of these three phenomena resulting in the direct upconversion of three FW photons, they all return very similar TH emission patterns, with a polarization parallel to the electric field of the impinging FW pump. For this reason, in the following we just describe the process in Fig. 3a as produced by an effective bulk $\chi^{(3)}$ term. We attribute this behavior, shared by all the three considered mechanisms, to the fact that our antenna geometry is characterized by a strong dipolar mode at the fundamental frequency $\omega$, while it lacks well-defined resonances at $3\omega$. Therefore, the THG resulting from the aforementioned processes is reminiscent of the large fields associated with the dipole induced at the FW, which points perpendicular to the nanorod long axis.

Concurrently, we described the emitted SHG (see Fig. 3b) by considering the nonlinear surface currents at the second harmonic (SH), $\mathbf{J}_{2\omega} = 2i\omega \mathbf{P}_{2\omega}$, with $\mathbf{P}_{2\omega} = \chi^{(2)}_{\perp\perp\perp} E^2_{\omega,\perp} \hat{\mathbf{n}}$ (where $\hat{\mathbf{n}}$ is the normal to the metal surface).[24,35,39-43] Other components of the $\chi^{(2)}$ tensor have been considered, but we found that they do not provide significant corrections. To account for the experimental conditions, the electric far field generated by each phenomenon was projected over the polarizer axis and integrated over a collection angle corresponding to the numerical aperture of the objective. We have also evaluated the non-local and gradient field effects described in ref. 38, which show the same qualitative SH emission polarization as the one obtained from the $\chi^{(2)}$-mediated SHG stemming from the surface. Both mechanisms, in fact, provide an SH emission polarized perpendicular to the FW, which we attribute to the fact that the nanostructures display a dipolar resonance at $2\omega$, with the induced SH dipole pointing parallel to the nanorod long axis. Therefore, in the following we just describe the SHG process as due to an effective surface $\chi^{(2)}_{\perp\perp\perp}$ term. Figures 3d and 3e show the calculated polar plots for the effective $\chi^{(3)}$-mediated THG and surface $\chi^{(2)}$-mediated SHG emission for both antennas R and V. The intensities in each plot are normalized to that corresponding to maximum emission, to highlight the relative changes in all channels. These two plots represent a situation where the $\chi^{(3)}$-mediated THG and the $\chi^{(2)}$-mediated SHG are completely uncoupled.

A ready comparison with the experimental data shows that the polarization behavior of the SHG (Fig. 2e) is perfectly reproduced by a model based on an effective surface $\chi^{(2)}_{\perp\perp\perp}$ process (Fig. 3e). Moreover, the hierarchy of the SHG intensity is also qualitatively reproduced, showing the most efficient SHG for the better tuned antenna R and a much smaller emission for antenna V. Conversely, the polarization behavior of the THG emitted by the investigated nanoantennas (Fig. 2d) cannot be reconstructed with an effective $\chi^{(3)}$ model, which should display a polarized emission parallel to the FW electric field (Fig. 3d) and with a degree of linear polarization higher than that of the SHG (Fig. 3e).

The behavior of the numerical results can be understood in terms of the symmetry properties of the electromagnetic mode that is excited at the FW. An important point to be stressed is that the nanorod in Antenna R is tailored to have the lowest-order resonance close to the SH frequency. Therefore, at the FW, the field inside the nanorod is negligible compared to the one inside the V-shaped structure, which, in turn, is not significantly affected by the presence of the rod (see Figs. 4a-c). This implies that the FW field distribution inside the antenna is symmetric, as highlighted by Figs. 4a-c. Therefore, considering only the FW fields in the model describing the upconversion process, the direction of the polarization axis of SHG and THG are model-independent and obey only parity-conservation rules. More specifically, for two-photon upconversion (SHG) the emission parity should be even, resulting in a polarization axis in the reflection symmetry plane of the V-shaped structure, while, for three photon upconversion (THG), the emission should have the same parity of the mode excited by the FW illumination, with the same polarization axis ($y$ axis in Fig. 1). These expectations are indeed confirmed by our simulations of both SHG and $\chi^{(3)}$-mediated THG and are corroborated be the experimental data obtained on isolated V-shapes, where the reflection symmetry of the structure is not spoiled by the presence of the nanorod (see also the Supplementary Information, Fig. S3).

As demonstrated by the maps in Figs. 4d-f, the reflection symmetry of the field distribution inside the antenna is broken at the SH frequency, at which the nanorod fundamental mode can be effectively excited, producing sizable fields. To reproduce the symmetry-breaking associated with the tilt of the THG polarization axis, we thus need to consider a significant contribution of the SH fields in the upconversion mechanisms. For this reason, we include in our model a cascaded second-order mechanism, corresponding to a SFG process where a third-harmonic photon is generated by the coherent build-up of a FW photon and a SH photon (see the scheme in Fig. 3c). Such mechanism, which is commonly observed in non-centrosymmetric bulk crystals, eventually adds up to the THG from pure $\chi^{(3)}$-related currents and, if properly optimized, allows boosting the THG efficiency in these media[44-48] and even in mesoscopic non-centrosymmetric systems.[49,50] The presence of cascaded effects in THG seeded by SFG in plasmonic nanoantennas has been neglected in the literature thus far because of the large difference often reported between the SHG and THG emission yields. However, sizeable SFG has been recently reported for some specific plasmonic nanostructures[15] and the optimization of SFG through intrapulse phase engineering has been exploited to maximize SHG in plasmonic nanoantennas.[47] Here we tentatively consider a surface $\chi^{(2)}_{\perp\perp\perp}$ term for SFG equivalent to that of SHG and, assuming both pump fields (FW and SH) are undepleted, we model THG as the cascade of a SHG and a SFG process described by nonlinear surface currents, $\mathbf{J}_{3\omega} = 2i\omega\,\chi^{(2)}_{\perp\perp\perp} E_{\omega,\perp} E_{2\omega,\perp}\hat{\mathbf{n}}$.

Figure 3f displays the purely cascaded THG emission for Antenna R and Antenna V evaluated from numerical simulations. While, as expected from parity conservation, the cascaded THG intensity calculated for Antenna V is symmetric with respect to the polarization of the impinging FW field (indicated by the double headed arrow in Fig. 3d), the main polarization axis of the cascaded THG emission from Antenna R is strongly tilted. Moreover, the plots in Fig. 3f show a shallower degree of linear polarization for Antenna R with respect to Antenna V. These features provide a rationale to justify all the experimental observations that a model based only on an effective $\chi^{(3)}$ is not able to reproduce. A significant contribution of a cascaded THG process would in fact account for (i) the tilts we systematically observe in the polarization axis of the total THG emission, (ii) the large variability of the tilt angles for different antenna geometries, and (iii) the depolarization with respect to the pump polarization state. For these reasons, we believe that the peculiar polarization state of the THG emission constitutes an undisputable evidence for the presence of the cascaded effect contributing to the THG from our nanostructures. We stress that this conclusion does not depend on the particular model employed to describe the upconversion phenomena in our plasmonic nanoantennas, but is unambiguously supported by considerations based on parity conservation alone: as stated above, the THG main polarization axis could differ from the one of the FW illumination only in the case symmetry-breaking SH fields inside the nanostructures play a significant role in determining their third-order nonlinear response.

A quantitative numerical estimate of the overall THG yield, accounting for both the $\chi^{(3)}$-mediated and the $\chi^{(2)}$-cascaded processes, is hindered by the intrinsic uncertainty associated with the values of the second-order and third-order effective nonlinear coefficients of gold. While recent theoretical models for $\chi^{(2)}_{\perp\perp\perp}$ [39-41] were shown to be effective in reproducing several experimental findings[41-43], the data reported in the literature for the $\chi^{(3)}$ value of gold are still subject to a large variability, due to the role played by the nonlinear incoherent effects and to the dependence of $\chi^{(3)}$ on the resonant conditions as well as on the experimental configurations.[25-29,31] Moreover, the different mechanisms contributing to $\chi^{(3)}$-mediated and $\chi^{(2)}$-cascaded THG phenomena introduce further uncertainty in a quantitative comparison between the intensities of these two contributions. Finally, while SHG is described by the nonlinear susceptibility $\chi^{(2)}_{\perp\perp\perp}(\omega,\omega)$, the SFG process leading to $\chi^{(2)}$-cascaded THG rather depends on a nonlinear susceptibility of the form $\chi^{(2)}_{\perp\perp\perp}(\omega,2\omega)$. Considering the strong dispersion of the nonlinear susceptibilities of metals, this is another source of uncertainties. Despite all these limitations, our numerical model is able to reproduce the presence of a tilted THG polarization axis for Antenna R, while, as also expected from symmetry considerations, no tilt is present for Antenna V.

The qualitative matching between experiments and simulations indicates that cascaded effects need to be included to model the THG, in striking contrast with earlier findings reporting negligible interaction between SHG and THG in plasmonic nanoantennas.[52] We also verified possible effects of the pump fluency on the cascaded THG by acquiring power-dependent emission polar plots (see Supplementary Information, Fig. S4). As expected in an undepleted pump regime, the power-dependent polar plots for both THG and SHG do not display any significant modification in the polarized emission as a function of the pump fluency.

A scenario where $\chi^{(2)}$-mediated processes at the nanoscale provide a sizeable emission compared to $\chi^{(3)}$-mediated THG qualitatively requires similar values of $\chi^{(3)}$ and $\left[\chi^{(2)}_{\perp\perp\perp}\right]^2$. At a first glance, this might appear incompatible with the values determined for the nonlinear coefficients $\gamma_{THG}$ – proportional to $[\chi^{(3)}]^2$ – and $\gamma_{SHG}$ – proportional to $\left[\chi^{(2)}_{\perp\perp\perp}\right]^2$ – since $\sqrt{\gamma_{THG}} \gg \gamma_{SHG}$. However, one should bear in mind that in all plasmonic nanostructures the SH dipoles radiate in the far field with an efficiency that is much smaller than the one of the TH dipoles. Due to parity selection rules, in fact, low radiation efficiencies are often attained in the far field for SHG, because of the out-of-phase oscillation of the SH dipoles at the surface of the structure. This leads to a complete suppression of the electric dipole emission mode of SH radiation in centrosymmetric nanostructures, while this cancellation is only partially relieved in systems like our nanoantennas, where the inversion symmetry is broken.[53] In both cases, the SH radiation associated with the local oscillating dipoles in the nanostructure destructively interfere in the far field, resulting in small $\gamma_{SHG}$ values even for a high effective $\chi^{(2)}_{\perp\perp\perp}$. Conversely, parity conservation in THG allows for the in-phase oscillation of dipoles either generated within the volume by third-order nonlinearities or at the surface by cascaded second-order nonlinear processes, resulting in a high radiation efficiency (and a high $\gamma_{THG}$ value) for both phenomena. A remarkable consequence of this conclusion is that significant fingerprints associated with the $\chi^{(2)}$-cascaded THG process can be expected even in plasmonic nanostructures characterized by low SHG efficiencies in the far field, which further corroborates the importance of our findings.

In conclusion, by studying the polarization state of the nonlinear emission in non-centrosymmetric gold antennas, we demonstrated that a $\chi^{(2)}$-cascaded process significantly contributes to THG. We underline that this conclusion does not depend on the model employed to describe second- and third-order optical nonlinearities but can be drawn only from the fact that parity conservation is broken only by SH fields. Such an upconversion mechanism based on a cascaded THG is particularly sensitive to the antenna geometry, since it stems from a sizeable SFG process[15] between the intense SHG and the pump photons. This result sheds new light on the processes behind THG in plasmonic nanoantennas and allows envisaging the possibility of further enhancing this phenomenon in the near infrared, where the influence of interband transitions becomes negligible. The ability to tailor and optimize the SFG process in plasmonic nanoantennas also offers new degrees of freedom for the optimization and integration of multiple optical functionalities into complex photonic devices and metasurfaces.[6] This represents a key step towards the realization of nanoscale and all-optical sensing devices for ultrafast switching and information processing in both classical and quantum optical systems at telecom wavelengths.

AUTHOR INFORMATION

**Corresponding Author**

*michele.celebrano@polimi.it

**Author Contributions**

All authors contributed to writing the manuscript and they have all approved the final version.

‡These authors contributed equally. (match statement to author names with a symbol)

ACKNOWLEDGMENT

The authors would like to thank G. Cerullo for fruitful discussions. This work has been carried out in the framework of two COST Actions: COST MP1403 Nanoscale Quantum Optics and COST MP1302 Nanospectroscopy.

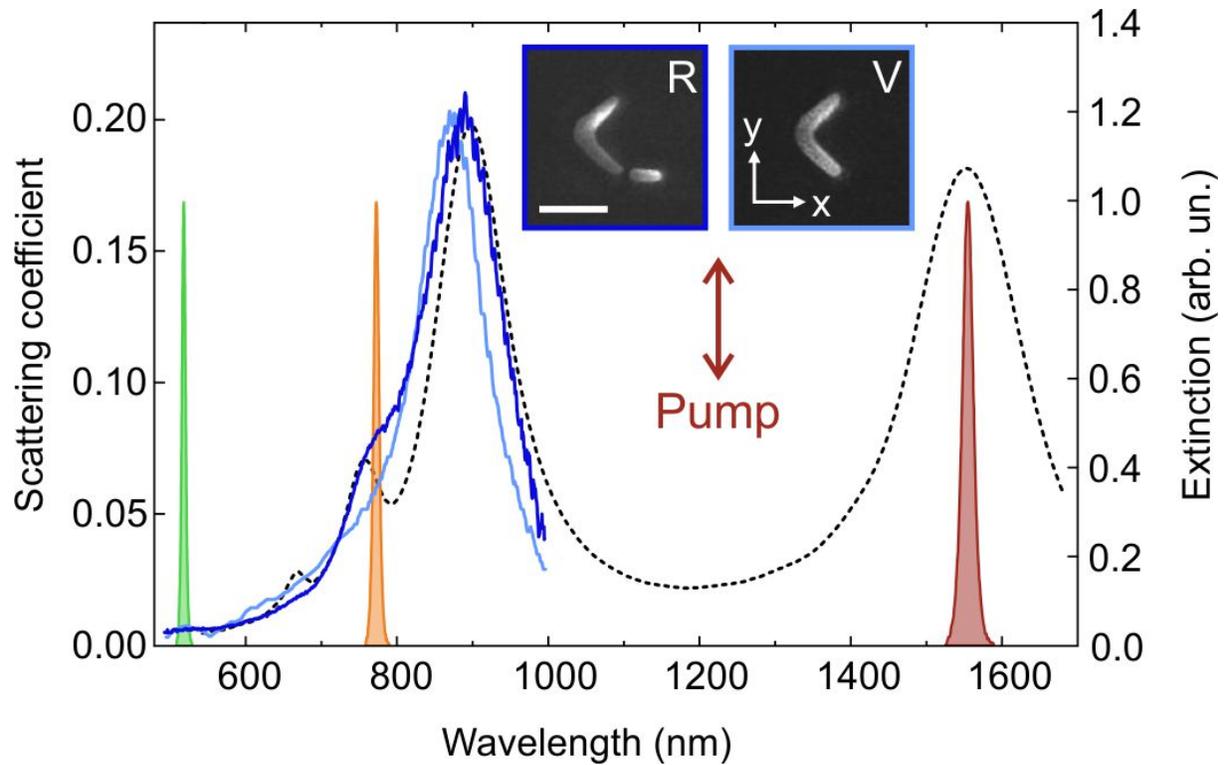

**Figure 1.** Whitelight spectra of the two investigated nanoantenna geometries: doubly resonant Antenna R (dark blue line) and symmetric Antenna V (light blue line). The dashed line is a finite-difference time-domain simulation of the scattering spectrum of Antenna R. The dark red, orange, and green filled curves indicate the hypotetic FW, SHG and THG lines. Insets: SEM images of Antenna R (dark blue) and Antenna V (light blue). The arm length of the V-shaped structure of the nanoantenna is about 160 nm for both antennas, while the rod length in Antenna R is 95 nm. The horizontal ruler corresponds to a length of 200 nm. The dark red double headed arrow represents the pump FW laser polarization with respect to the nanostructures in the insets.

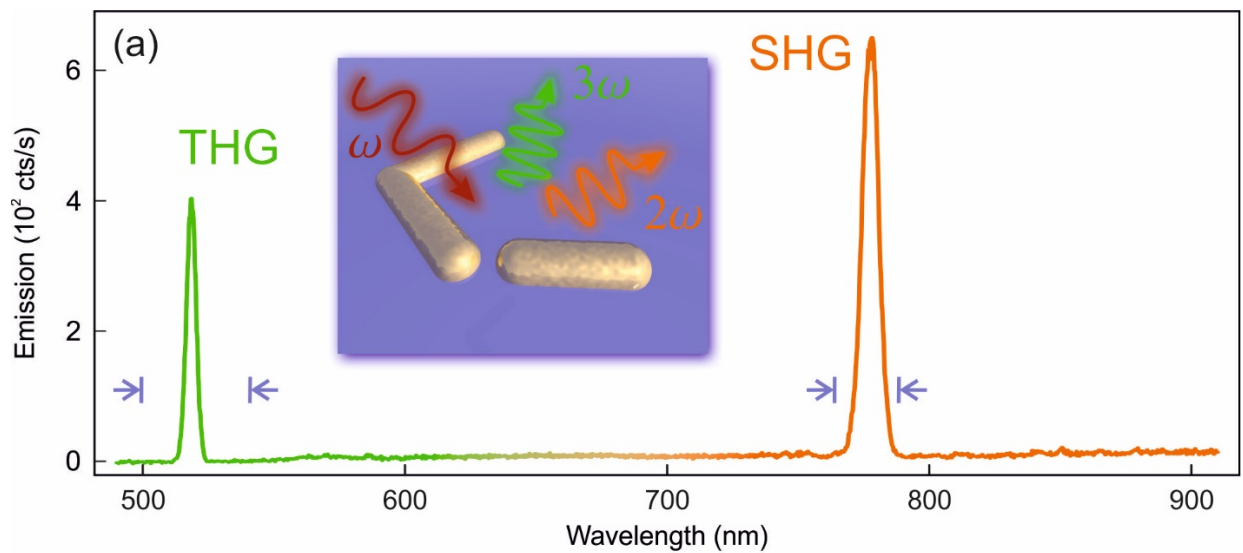
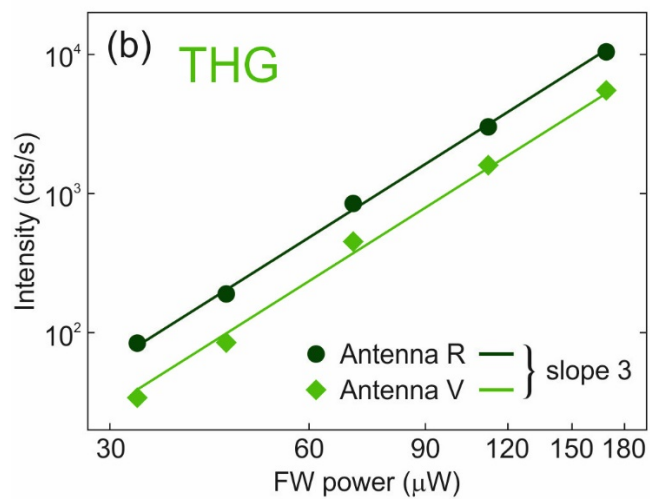
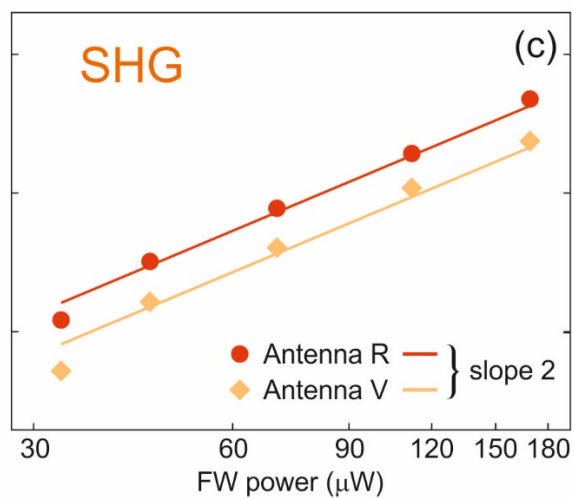
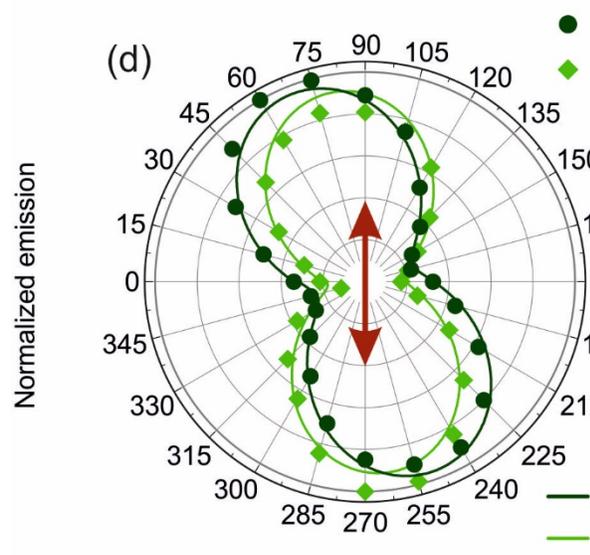
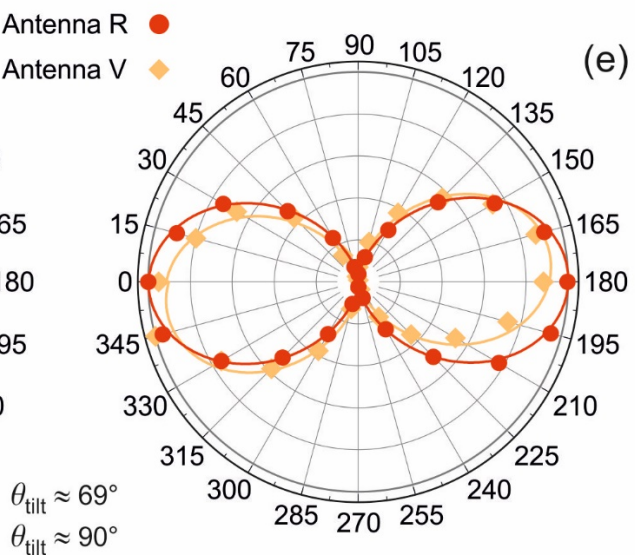

**Figure 2.** (a) Nonlinear emission spectrum of the double resonant Antenna R. The green side represents the visible part of the spectrum, while the orange side represents the NIR part. The light blue arrows indicate the rising and falling edges of the band-pass filters employed for the analysis of THG and SHG. The two filters are centered at 520 nm and 775 nm with bandwidths of 40 nm and 25 nm, respectively. The THG emission is peaked at 518 nm (linewidth ≈ 4 nm), while the SHG line is centered at 777 nm (linewidth ≈ 7 nm). Inset: picture of the nanoantenna with the color-coded harmonics involved in the processes. (b) THG (green) and (c) SHG (orange) intensities collected from Antenna R and Antenna V as a function of the input power. The solid lines represent linear fits to the experimental points. (d, e) Polar plots of the normalized intensity emitted by THG (d) and SHG (e) for the two antennas. The polar plot angle $\theta$ is defined with respect to the horizontal $x$ axis shown in Fig. 1. Color codes are the same as for panels (b) and (c). Solid lines correspond to the best fit performed with the function $A' \cos^2(\theta - \theta_0) + B'$ for the SHG plots, or the function $A'' \cos^2(\theta - \theta_{tilt} - \theta_0) + B''$ for THG plots. The double-headed dark red arrow represents the direction of the pump linear polarization.

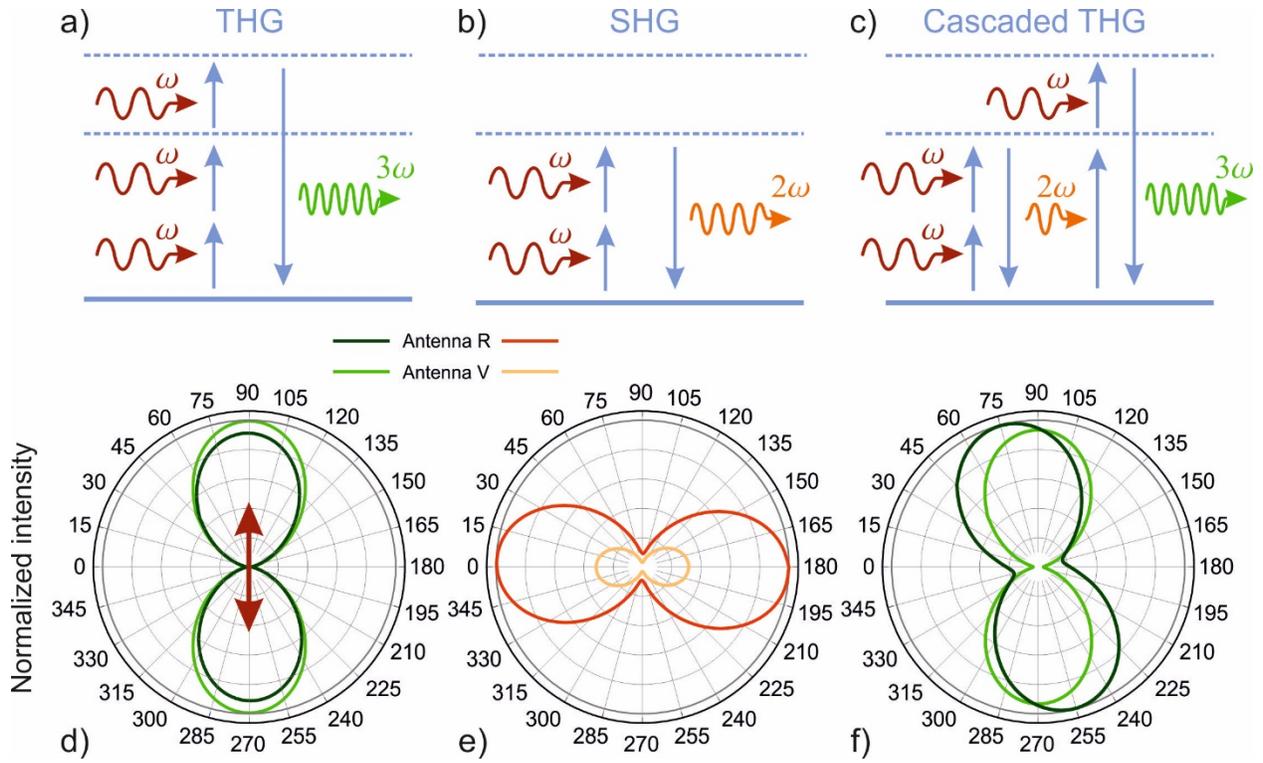

**Figure 3.** (a-c) Conversion schemes of the (a) THG, (b) SHG and (c) cascaded THG processes that are mediated by the nanoantenna. (d-f) Simulated emission polar plot for (d) purely $\chi^{(3)}$-mediated THG, (e) $\chi^{(2)}$-mediated SHG, and (f) cascaded THG. The color code is the same as in Fig. 2. In each plot, the intensities are normalized to that of the most intense emission. The impinging light polarization is indicated by the double arrow in panel (d) and is oriented perpendicular to the symmetry plane of the V-shape structure ($y$ axis in Fig. 1).

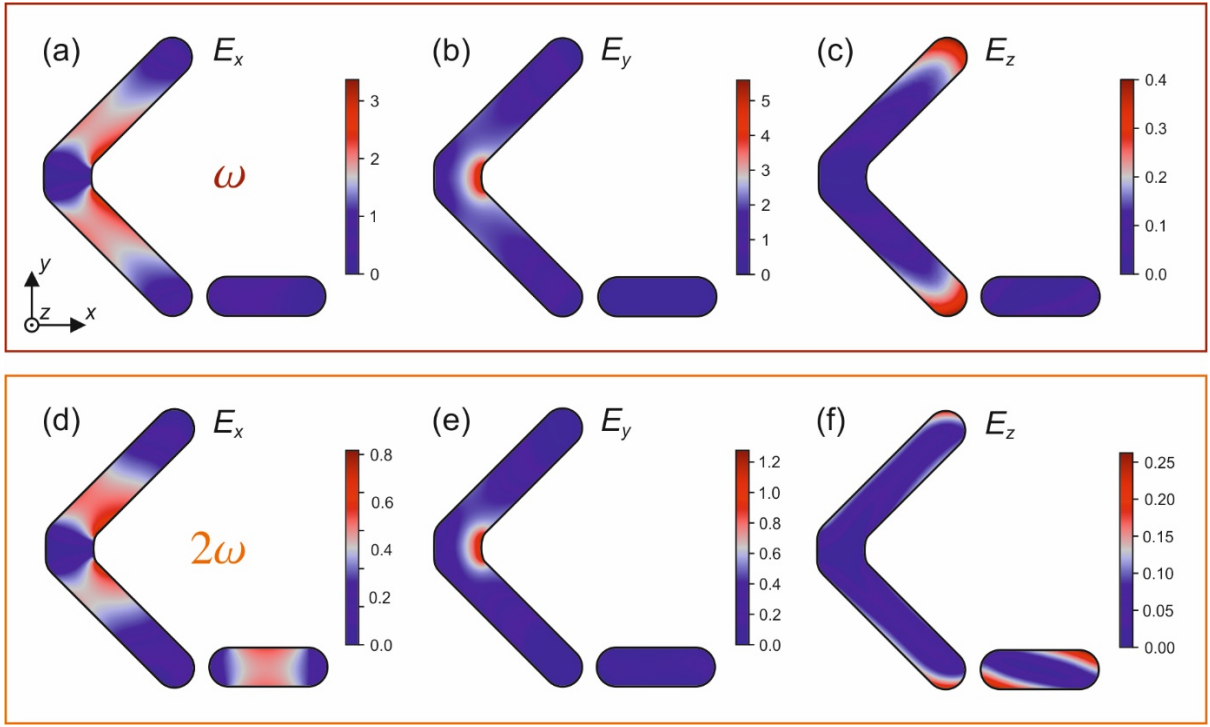

**Figure 4.** Numerical simulations reporting the amplitude of the *x*, *y*, and *z* components of the electric field inside Antenna R, excited by a plane electromagnetic wave oscillating either at the fundamental frequency $\omega$ and linear polarization along the *y* axis (a-c) or at *2ω* and linear polarization along the *x* axis (d-e). The color scales are expressed in units of the amplitude of the electric field of the impinging waves at either $\omega$ or $2\omega$, respectively. The maps refer to the electric field amplitudes calculated in the median plane crossing the antenna at half thickness.

REFERENCES

1. De Martini, F.; Shen, Y. R. Nonlinear Excitation of Surface Polaritons. *Phys. Rev. Lett.* **1976**, 36, 216-219.

2. Chen, C. K.; de Castro, A. R. B.; Shen, Y. R. Surface-Enhanced Second-Harmonic Generation. *Phys. Rev. Lett.* **1981**, 46, 145-148.

3. Smolyaninov, I. I.; Zayats, A. V; Davis, C. C. Near-field second harmonic generation from a rough metal surface. *Phys. Rev. B* **1997**, 56, 9290–9293.

4. Zayats, A. V., Kalkbrenner, T., Sandoghdar, V. & Mlynek, J. Second-harmonic generation from individual surface defects under local excitation. *Phys. Rev. B* **2000**, 61, 4545–4548.

5. Kauranen, M.; Zayats, A. V. Nonlinear Plasmonics. *Nature Photon.* **2012**, 6, 737-748.

6. Butet, J.; Brevet, P. F.; Martin, O. J. F. Optical Second Harmonic Generation in Plasmonic Nanostructures: From Fundamental Principles to Advanced Applications. *ACS Nano* **2015**, 9, 10545–10562.

7. Li, G.; Zhang, S.; Zentgraf, T. Nonlinear photonic metasurfaces. *Nat. Rev. Mat.* **2017**, 2, 17010.

8. Maser, A.; Gmeiner, B.; Utikal, T.; Götzinger, S.; Sandoghdar, V. Few-photon coherent nonlinear optics with a single molecule. *Nat. Photon.* **2016**, 10, 450–453.

9. Boyd, R. W. *Nonlinear Optics*, 3$^{rd}$ Ed.; Academic Press: Cambridge, **2008**; pp 79-84.

10. Novotny, L.; van Hulst, N. F. Antennas for light. *Nat. Photon.* **2011**, 5, 83–90.

11. Biagioni, P.; Huang, J.-S.; Hecht, B. Nanoantennas for visible and infrared radiation. *Rep. Prog. Phys.* **2012**, 75, 024402.

12. Hanke, T.; Cesar, J.; Knittel, V.; Trügler, A.; Hohenester, U.; Leitenstorfer, A.; Bratschitsch, R. Efficient nonlinear light emission of single gold optical antennas driven by few-cycle near-infrared pulses. *Phys. Rev. Lett.* **2009** 103, 257404.

13. Danckwerts, M.; Novotny, L. Optical Frequency Mixing at Coupled Gold Nanoparticles. *Phys. Rev. Lett.* **2007**, 98, 026104.

14. Harutyunyan, H.; Volpe, G.; Quidant, R.; Novotny, L. Enhancing the Nonlinear Optical Response Using Multifrequency Gold-Nanowire Antennas. *Phys. Rev. Lett.* **2012**, 108, 217403.